\documentclass[pra, showpacs,amsmath,amssymb,twocolumn, 10pt]{revtex4-1}

\usepackage{amsthm}
\usepackage{dcolumn}
\usepackage{bm}
\usepackage{graphicx}
\usepackage{color}

\begin{document}
\newtheorem{question}{Question}
\newtheorem{corollary}{Corollary}
\newtheorem{definition}{Definition}
\newtheorem{example}{Example}
\newtheorem{lemma}{Lemma}
\newtheorem{proposition}{Proposition}
\newtheorem{statement}{Statement}
\newtheorem{theorem}{Theorem}
\newtheorem{property}{Property}
\newtheorem{fact}{Fact}
\newtheorem{conjecture}{Conjecture}

\newcommand{\bra}[1]{\langle #1|}
\newcommand{\ket}[1]{|#1\rangle}
\newcommand{\braket}[3]{\langle #1|#2|#3\rangle}
\newcommand{\ip}[2]{\langle #1|#2\rangle}
\newcommand{\op}[2]{|#1\rangle \langle #2|}

\newcommand{\tr}{{\mathrm{tr}}}
\newcommand{\dt}{{{\Delta}t}}
\newcommand{\me}{{\mathrm{e}}}
\newcommand{\mi}{{\mathrm{i}}}
\newcommand {\A } {{\mathcal{A}}}
\newcommand {\E } {{\mathcal{E}}}
\newcommand {\U } {{\mathcal{U}}}
\newcommand {\D } {{\mathcal{D}}}
\newcommand {\M} {{\mathcal{M}}}
\newcommand {\T } {{\mathcal{T}}}
\newcommand {\I } {{\mathcal{I}}}
\renewcommand{\b}{\mathcal{B}}
\newcommand{\hs}{\mathcal{H}}

\title{Debugging Quantum Processes Using Monitoring Measurements}
\author{Yangjia Li$^{1,2}$}
\email{liyj04@mails.tsinghua.edu.cn}
\author{Mingsheng Ying$^{2,1}$}
\email{Mingsheng.Ying@uts.edu.au}

\affiliation{$^1$State Key Laboratory of Intelligent Technology and
Systems, Tsinghua National Laboratory for Information Science and
Technology, Department of Computer Science and Technology,
Tsinghua University, Beijing 100084, China\\
$^2$Centre for Quantum Computation and Intelligent Systems (QCIS),
Faculty of Engineering and Information Technology, University of
Technology, Sydney, NSW 2007, Australia}

\date{\today}

\begin{abstract}
Since observation on a quantum system may cause the system state collapse, it is usually hard to find a way to monitor a quantum process, which is a quantum system that continuously evolves. We propose a protocol that can debug a quantum process by monitoring, but not disturb the evolution of the system. This protocol consists of an error detector and a debugging strategy. The detector is a projection operator that is orthogonal to the anticipated system state at a sequence of time points, and the strategy is used to specify these time points. As an example, we show how to debug the computational process of quantum search using this protocol. By applying the Skolem--Mahler--Lech theorem in algebraic number theory, we find an algorithm to construct all of the debugging  protocols for quantum processes of time independent Hamiltonians.
\end{abstract}


\maketitle
\section{Introduction}
A major problem in physical implementation of quantum computation is that errors are usually unavoidable in practical situation. To protect the computing process against errors, the method of fault-tolerant quantum computation~\cite{Sho96,Kit97} has been introduced and developed in the last eighteen years. By employing many techniques of quantum error correction~\cite{Sho95,Ste96,NC00}, this method often leads to results in a form of threshold theorem~\cite{Kit97,KLZ98}: A quantum computer can be successfully implemented with high probability if each component of the system only fails with probability less than a threshold. The fault-tolerant quantum computation is usually used when the errors are caused by environment noises. The threshold condition is possibly satisfied in this case, as the interaction between the quantum system and the environment may be reduced by other physical techniques, such as~\cite{ZSBS14}.

In the present paper, we propose a so-called ``debugging'' method to deal with another type of errors that are not caused by environment noises but by ``bugs'', which mean unknown defects in the physical system itself. The prior techniques for fault-tolerant computation would generally become ineffective for such errors, since the error threshold is mostly broken. For example, suppose a Hadamard gate is by mistake used as a NOT gate in a quantum computer, then this small defect will greatly change the computing result in most cases. A more reasonable strategy here is to find the nature and exact position of this defect, and then repair it. To this end, quantum measurements should be applied to monitoring the computing process so that errors can be detected as soon as possible after the component with bugs being executed. Remarkably, a debugging method like this plays an indispensable role and attracts intense studies~\cite{Mye79} in the implementation of classical computing systems.

Unfortunately, due to the fundamental difference between the physical behavior of quantum measurements and that of classical ones, debugging for quantum systems is much more difficult than for classical systems, and thus classical debugging method does not work in the quantum scenario. Specifically, in quantum mechanics, observation of a quantum system would make the system state collapse. This interaction between observing apparatus and quantum systems on the one hand allows quantum measurements to drive target systems as quantum operations~\cite{DFY09,NC00}, in applications like teleportation~\cite{BBC+93}, entanglement distillation~\cite{BBP+96}, control of quantum systems~\cite{AT12}, and one-way quantum computing~\cite{RB01}; but on the other hand, it makes many tasks much harder than in the classical world, particularly when quantum measurements are used to extract (classical) information of given systems. The well known indistinguishability between nonorthogonal states can somehow be seen as a simple example. The quantum debugging task considered here is actually another instance, where measurements monitor the system state for possible errors. This can be easily done for a classical process, because the trajectory of a classical system is unchanged by measurements. However, a problem in monitoring a quantum process is that once the system had been measured, the system state may be disturbed and then be useless for further processing. This problem has been demonstrated to be very serious in the quantum Zeno effect~\cite{MS77}, that a quantum process can be completely obstructed by continuing measuring. Therefore, similar tasks are usually achieved by quantum tomography techniques~\cite{PCZ97,NC00} in the literature, in which the system state is measured only once to keep the outcome faithful, but instead, a large number of copies of the process are required.

\begin{figure}
  \includegraphics{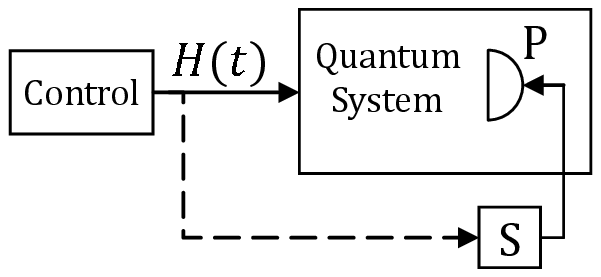}
\caption{The classical control information about $H(t)$ is sending to $S$ during the execution of the process. Then at any time $t$, $S$ can decide whether or not $P\ket{\psi_t}=0$ according to the control history. And if is it will drive $P$ to detect possible errors.}
\label{fig:1}
\end{figure}

In the debugging method proposed in this paper, quantum measurements are used in a different way: they are constantly taken to monitor a quantum process but without disturbances on the system state, until an error is detected. A basic scheme is described as follows. Consider a quantum system that is established to run some computing process. It is designed to be in state $\ket{\psi_0}$ initially, and then evolves under the controlled Hamiltonian $H(t)$. In this way, the trajectory $\{\ket{\psi_t}\}$ of the system state would be as anticipated. The time for the whole process is considered to be infinite, as it is usually much longer than the time for a single component (like a gate) acting. Now suppose a bug of the system will be involved in the process at time $t'$, then the system Hamiltonian will not truly be $H(t)$ for $t\geq t'$ in the practical execution. This causes errors in system state, so we write $\rho_t$ for the density operator of the actual state at time $t$.  To debug the process, we need to find a projection operator $P\neq0$ of the system and a sequence of time points $t_1,t_2,\cdots\ (t_n\rightarrow\infty)$, such that $P\ket{\psi_{t_n}}=0$ for all $n$.
The condition of $P\ket{\psi_t}=0$ means that nothing can be detected by $P$ if the system state is $\ket{\psi_t}$ as anticipated.
We monitor the process at time $t_1,t_2,\cdots$, using a measurement apparatus formalized by $P$. This measurement is called a monitoring measurement. Then with probability $\tr(P\rho_{t_n})$ the error state would be detected at time $t_n$. If it really happens, then an error is detected in the state. In this case, $t'$ is more likely in $[t_{n-1},t_n]$ and the relevant components should be carefully checked. Practically, the time points $t_1,t_2,\cdots$ are determined by a classical program $S$. Then the debugging protocol is visualized as FIG.~\ref{fig:1}. Obviously, the key step of debugging a process is to find the required projection operator $P$. The condition of $P\ket{\psi_t}=0$ guarantees that the anticipated process is not disturbed by $P$. On the other hand, it implies that the protocol is conclusive; i.e., no errors would be reported when the process runs correctly.

The aim of this paper is to develop the debugging method for quantum systems outlined above. The paper is organized as follows. In Sec.~\ref{sec:proto}, we first consider an example debugging protocol for quantum search algorithm. After that, we propose a general debugging scheme and show that it can be reduced to a simpler scheme described as in FIG~\ref{fig:1}. Then we formally define this simplified debugging protocols in the case of discrete time evolution. In Sec.~\ref{sec:resu}, we completely solve the debugging problem for quantum processes with time independent Hamiltonians. More precisely, we find an algorithm to construct all of the debugging protocols for this kind of quantum processes by employing the celebrated Skolem--Mahler--Lech theorem. A brief conclusion is drawn in Sec.~\ref{sec:conclu}.

\section{Debugging Protocols}\label{sec:proto}
\subsection{An Example}
To show how can the debugging method be truly applied, let us first consider a simple example --- debugging for the computational process of quantum search~\cite{Gro96}. Here, we adopt the description of the Grover algorithm in \cite{NC00}. The quantum computer consists of $n$ qubits with $\ket{0}^{\otimes n}$ as the initial state (for simplicity, we omit the auxiliary qubits of the oracle). A black box oracle $O$ of form $$O=I_2^{\otimes n}-2\op{x}{x}$$ is provided as input, where $x\in\{0,1,\cdots,2^n-1\}$ is the index we want to find. The computer first applies $H_2^{\otimes n}$, and then successively applies the Grover iteration $$G=(2\op{\psi_0}{\psi_0}-I_2^{\otimes n})O$$ for $O(\sqrt{2^n})$ times, where $$\ket{\psi_0}=\sum_{k=0}^{2^n-1}\ket{k}/\sqrt{2^n}.$$ At last, $x$ can be obtained with probability $O(1)$ by measurement in the computational basis $\{|0\rangle,|1\rangle\}$ on each qubit. Here, we use $I_2$ and $H_2$ to denote the identity and Hadamard gates, respectively.

To debug this process, we note that the system state immediately after each Grover iteration should be always in the two-dimensional subspace $\mathrm{span}\{\ket{x},\ket{\xi}\}$, where $\ket{\xi}=\sum_{k\neq x}\ket{k}/\sqrt{2^n-1}$. So, we can use a measurement apparatus formalized by $P=I_2^{\otimes n}-\op{x}{x}-\op{\xi}{\xi}$ to detect errors. The protocol is as follows: randomly choose an integer $x$ and provide the corresponding oracle at the beginning, and then execute the algorithm. Immediately after each Grover iteration $G$, take the monitoring measurement $P$ to detect errors. If an error system state is detected at some time point, then the debugging protocol stops the process and reports this error. Now we particularly discuss the following two kinds of bugs: \begin{enumerate}\item The system was not initialized. We write $\rho$ for the density operator of the initial system state and write $f$ for the fidelity of $\rho$ and $\ket{0}^{\otimes n}$. Then it is easy to verify that with probability $(2^n-2)(1-f^2)/(2^n-1)$ an error can be detected just by the first measurement of $P$.
\item The Grover iterator was implemented with some bugs, so it is not $G$ but some unitary operator $G'$. In most cases, the two subspaces $\mathrm{span}\{G'\ket{x},G'\ket{\xi}\}$ and $\mathrm{span}\{G\ket{x},G\ket{\xi}\}=\mathrm{span}\{\ket{x},\ket{\xi}\}$ have no common state. So $|\braket{x}{G'}{\psi}|^2+|\braket{\xi}{G'}{\psi}|^2<1$ for all $\ket{\psi}\in \mathrm{span}\{\ket{x},\ket{\xi}\}$. We write $q$ for the maximal value of all $|\braket{x}{G'}{\psi}|^2+|\braket{\xi}{G'}{\psi}|^2$. Then at each measurement of $P$, an error will be detected with a positive probability at least $1-q>0$.
\end{enumerate}
Two advantages of the quantum debugging method are demonstrated in this example: (1) An error may be detected soon after the bugs involved. So, the process can be just partly executed and a lot of time would be saved; (2) A single execution of the process is usually sufficient to detect an error, whereas a large number of copies of the process are needed in other approaches.

\subsection{A General Debugging Protocol}
We now consider a general scheme of debugging protocols for quantum processes, in which quantum measurements are in the most general form, and different measurements can be used at different time points to detect errors. First, we impose a \emph{compatibility} constraint to each monitoring measurement, such that the target system state keeps unchanged under its action. Formally, the compatibility can be stated as follows: let $\ket{\psi}$ be a state and $\M=\{M_1,M_2,\cdots,M_k\}$ be a measurement. We say that $M$ is compatible with $\ket{\psi}$ if for all $i$, $M_i\ket{\psi}$ is essentially the same as $\ket{\psi}$ or vanish; that is, $\ket{\psi}$ is an eigenstate of every measurement operator of $\M$:
\begin{equation}\label{equ:consistentness}\forall i\ \exists \lambda_i\ {\rm s.t}\ M_i\ket{\psi}=\lambda_i\ket{\psi}.\end{equation}
In fact, this constraint simulates the physical behavior of a classical measurement: The states of a classical system can be thought of as an orthonormal basis $\{\ket{i}\}_{i=1}^k$, and we consider a classical measurement $\M=\{M_1,...,M_k\}$ with $M_i=\op{i}{i}$. Then the compatibility is automatically satisfied: $M_i\ket{j}=\delta_{ij}\ket{j}$ for each $i$ and $j$.

A general protocol for debugging a quantum process using monitoring quantum measurements consists of three steps: \begin{enumerate}\item Set a sequence of breakpoints at time $t_1,t_2,\cdots$ during the process; \item Execute the process, and at each breakpoint of time $t_n$, insert a measurement $\M_{t_n}=\{M_{t_n1},M_{t_n2},\cdots,M_{t_nk_n}\}$ that is compatible with the anticipated system state $\ket{\psi_{t_n}}$. We write $E(M_{t_n})=\{i|M_{t_ni}\ket{\psi_{t_n}}=0\}$ for the outcomes $i$ that should not occur at time $t_n$ if the process behaves as anticipated; \item An error is detected if the measurement outcome at $t_n$ is some element $i\in E(M_{t_n})$. In this case we stop the execution and report $i$ and $t_n$ to specify the error type and the error position, respectively.\end{enumerate}

We can actually simplify this general debugging scheme without loss of generality. First, we show that at each breakpoint of time $t$, the general quantum measurement $\M_t=\{M_{ti}\}$ can be replaced with the two-outcome POVM $\{I-E_t, E_t\}$, where $E_t=\sum_{i\in E(\M_t)}M_{ti}^\dagger{M_{ti}}$ is used to indicate errors and $I-E_t$ indicates correctness. Here $I$ is the identity operator of the system. In fact, this POVM performs mostly the same as $\M_t$: They are both compatible with $\ket{\psi_t}$ and detect errors with the same probability $\tr(E_t\rho_t)$. Here we denote by $\rho_t$ the system state with errors. The only disadvantage of such replacement is that different error types $i$ in $E(\M_t)$ are not distinguished. However, this would not be a problem because after an error being detected, the type can be specified by further measurement. Moreover, it is even better to use the projective measurement $\{I-P_t,P_t\}$ with $P_t$ being the projection operator into the support of $E_t$. This is because the measurement also satisfies the compatibility, and it detect errors with probability $\tr(P_t\rho_t)\geq\tr(E_t\rho_t)$. Therefore, it suffices to detect errors using monitoring measurements formalized by projection operators $P_t$. We will call them error detectors in what follows.

Secondly, we assert that all of the error detectors $P_t$ should be chosen only from a finite set; otherwise, the protocol would be useless. The reason is that if infinitely many detectors are used, then to decide which one is chosen at a breakpoint, the amount of information we needed would become infinite. A specific instance is helpful to understand this situation: At each breakpoint of time $t$, we simply use $P_t=\psi_t^\perp$ as the error detector. Obviously, it is compatible with $\ket{\psi_t}$ and any error of this system state can be detected using it. However, to construct this detector we need the complete information of $\ket{\psi_t}$ by classical computation; namely, the debugging protocol requires a classical simulation of the quantum process, which is clearly unreasonable. So, the requirement of finiteness is crucial for effective debugging protocols. We write all the detectors as $P_1,P_2,\cdots,P_k$. Then there is a strategy $S$ for the protocol to call one of them at each breakpoint. Now we can divide the strategy $S$ into $k$ parts $S_1,S_2,\cdots,S_k$, where $S_i$ is a strategy that only call $P_i$ at corresponding breakpoints and keeps silent at the others. Then the original debugging protocol can be decomposed as $k$ protocols $(P_i,S_i) (i=1,2,\cdots,k)$, each of which monitors the process at a part of breakpoints. In particular, some of the protocols will constantly work at an infinite subsequence of the breakpoints.

Therefore, we only need to investigate the debugging protocols of such a form: it consists of an error detector $P$ and a strategy $S$; at a sequence of time points specified by $S$, $P$ is taken to detect possible errors of the system state. We note that this simplified protocol is exactly that visualised in FIG.~\ref{fig:1}. If all protocols in this scheme can be found for a given quantum process, then a general debugging task can be achieved by a simple combination of them, with certain further analysis about the detected errors.

\subsection{Discrete Time Evolution}
Since an error detector $P$ is discretely taken in the debugging described above, it is reasonable to consider the discrete time evolution of the system. Specifically, we assume that the compatibility constraint is only checked by strategy $S$ at given points $t_0,t_1,...$ of time. Then it suffices to considering the corresponding states $\ket{\psi_{t_0}},\ket{\psi_{t_1}},\cdots$, and the state transformations between them, which are formalized by unitary operations. In this way, the design of a quantum process can be depicted as $$\ket{\psi_{t_0}}\overset{U_{\alpha_1}}\rightarrow\ket{\psi_{t_1}}\overset{U_{\alpha_2}}\rightarrow\ket{\psi_{t_2}}\overset{U_{\alpha_3}}\rightarrow\cdots,$$
where $\ket{\psi_{t_n}}=U_{\alpha_n}\ket{\psi_{t_{n-1}}}$ for every $n\geq 1$, and $U_{\alpha_n}$ describes the evolution of the system from time $t_{n-1}$ to $t_n$. For realizability, we can assume that all of these unitary operators can be chosen from a finite set $\{U_1,...,U_m\}$. Then we have $\alpha_n\in\{1,...,m\}$ for every $n=1,2,...$. Obviously, a circuit model of quantum computation can be seen as a quantum process like this, where $U_1,...,U_m$ are the gates in the circuit. Quantum walk~\cite{ABN+01} can be considered as another example of quantum processes in this form.

Now we rigorously define the debugging protocol $(P,S)$ for quantum processes formulated by such a system. An error detector $P$ is a projection operator in the state Hilbert space $\hs$, and a strategy $S$ is a function that to each finite sequence $s=\alpha_1\alpha_2\cdots\alpha_n$ of indices in $\{1,...,m\}$, assigns a result of ``yes'' or ``no''. Intuitively, $S(s)=$``yes'' (resp. ``no'') means that $P$ is (resp. not) used to detect errors immediately after the execution of the action sequence $U_{\alpha_1},U_{\alpha_2},\cdots,U_{\alpha_n}$. For simplicity, we write $U_s=U_{\alpha_n}\cdots U_{\alpha_2}U_{\alpha_1}$ for the composition of the corresponding unitary actions. To warrant the protocol actually realizable, the following three conditions are necessary:
\begin{enumerate}
\item (Compatibility) $S(s)=$``yes'' implies $PU_s\ket{\psi_{t_0}}=0$.
\item (Computability) A classical algorithm can be found to compute $S$.
\item (Liveness) For any infinite sequence $\alpha_1\alpha_2\cdots$ of indices $1,...,m$ there are infinitely many $n$'s such that $S(\alpha_1\alpha_2\cdots\alpha_n)=$``yes''.
\end{enumerate} The first two conditions are easy to understand. The liveness comes from the fact that $P$ should constantly be applied in the process represented by $\alpha_1\alpha_2\cdots$, so that bugs involved at any time could be detected.

Based on the above definition of debugging protocol, a debugging problem can be formally stated as follows: \begin{itemize}\item Given an initial state $\ket{\psi_{t_0}}$ and a set of unitary operations $U_1,U_2,\cdots, U_m$ that describe the discrete-time evolution of a quantum process, how can we find all the protocols $(P,S)$ satisfying Compatibility, Computability and Liveness?\end{itemize}

\section{Debugging for Time-Independent Hamiltonians}\label{sec:resu}

\subsection{A Basic Theorem}

We now solve the debugging problem for the case where the designed Hamiltonian is time independent. Specifically, our solution consists of the following three steps: \begin{enumerate}\item We find a method to check whether or not a given projection operator $P$ can be used as an error detector; \item For each eligible $P$, we show that a strategy $S$ can be constructed as a periodic function; \item We present a procedure that can compute all the debugging protocols $(P,S)$ for any given process.\end{enumerate}

Let $\hs$ be the state Hilbert space of the system, and $H$ the system Hamiltonian which is time independent. To define debugging protocols $(P,S)$, we consider the discrete time evolution of the system between a sequence of time points $0,\dt,2\dt,\cdots$, where $\dt$ is a fixed period of time which can be appropriately chosen in practice. Then at time $n\dt$, the anticipated system state is $\ket{\psi_n}=U^n\ket{\psi_0}$, where $\ket{\psi_0}$ is the initial state and $U=\exp(-\mi H\dt/\hbar)$ is the unitary transformation of time evolution in a single period. As defined in Subsec.~\ref{sec:proto}-C, a debugging protocol for this system consists of an error detector $P$ which is an projection operator of $\hs$, and a strategy $S$ which is a function specifying (by assigning ``yes'') an infinite sequence of integers $i_1,i_2,\cdots$ such that $PU^{i_n}\ket{\psi_0}=0$ for all $n$. Our task is to find the detector $P$ and the strategy $S$.

Obviously, a necessary condition of $P$ being an error detector is that $PU^n\ket{\psi_0}=0$ for infinitely many $n$. To investigate how this condition can be satisfied, we need the following theorem:
\begin{theorem}\label{the:main}{\rm Let $\ket{\psi_0}$ be a vector, $U$ a unitary operator and $P$ a projection operator in a finite dimensional space $\hs$. If $Z=\{n|PU^n\ket{\psi_0}=0\}$ is an infinite set, then an arithmetic progression $\{pn+r|n=0,1,\cdots\}$ can be algorithmically found in $Z$.}
\end{theorem}

The proof of Theorem~\ref{the:main} is postponed to next subsection. Here we see how this theorem can be used in our investigation of a debugging protocol $(P,S)$. First, the infiniteness condition of $Z$ can be checked, as it is equivalent to the existence of the arithmetic progression. Second, this condition is not only necessary but also sufficient for $P$ being an error detector. In fact, if it holds for $P$, then we can construct a strategy $S$ as a periodic function that assigns ``yes'' to the integers $pn+r$, $n=0,1,\cdots$, and ``no'' to the others. Moreover, by making the arithmetic progression $\{pn+r|n=0,1,\cdots\}$ exist in $Z$, we have a procedure to compute all the error detectors $P$. Such a procedure will be carefully described in Subsec. \ref{sec:resu}-C based on the proof of the theorem.

\subsection{Proof of Theorem~\ref{the:main}}\label{proofs}
A key step in the proof of Theorem~\ref{the:main} is to explore the implication of the infiniteness of $Z$. For this purpose, we employ some techniques from the previous research on the famous Skolem's problem~\cite{Skolem34}. Consider a linear recurrent sequence $\{a_n\}_{n=0}^\infty$, which satisfies the linear recurrence relation:
\begin{equation}\label{equ:recurrence}
a_{n+d}=c_{d-1}a_{n+d-1}+c_{d-2}a_{n+d-2}+\cdots+c_0a_{n}
\end{equation}
for all $n\geq0$. Let $Z=\{n|a_n=0\}$ be the set of indices of null elements of $\{a_n\}$. A way relating the above linear recurrent sequence to the behavior of a quantum system is putting $a_n=\braket{\phi}{M^n}{\psi}$ for two quantum states $\ket{\phi},\ket{\psi}$ and a quantum operation $M$ of a $d$ dimensional quantum system. Remarkably, this technique has already been successfully used to solve several important problems in quantum information theory. For example, the condition $\braket{\phi}{M^n}{\psi}=0$ is interpreted as the acceptance condition of finite quantum automata in \cite{BJKP05} for $M$ being a unitary operator, and as the occurrence of specific quantum measurement outcomes in \cite{EMG12} for $M$ being a measurement operator, respectively. The decision problems considered in~\cite{BJKP05,EMG12} are similar to the Skolem's emptiness problem~\cite{HHHK05}. What we need in the proof of our result is the following~\cite{Lech52}:
\begin{theorem}[Skolem--Mahler--Lech]\label{the:SML}
{\rm In a field of characteristic 0, let a sequence $\{a_n\}_{n=0}^\infty$ satisfy a recurrence relation of form Eq.~(\ref{equ:recurrence}), then the set $Z$ of indices of null elements of this sequence is semi-linear, namely, is a union of a finite set and finitely many arithmetic progressions.}
\end{theorem}

To apply this theorem to our problem, we decompose $P=\sum\ket{\phi_i}\bra{\phi_i}$, where states $\ket{\phi_i}$ form an orthonormal basis of the image space of $P$. Let $\lambda^d-c_{d-1}\lambda^{d-1}-c_{d-2}\lambda^{d-2}-\cdots-c_0$ be the characteristic polynomial of $U$. Then for each $\ket{\phi_i}$, we can invoke Theorem~\ref{the:SML} for $a_n=\braket{\phi_i}{U^n}{\psi_0}$ and assert that the set $Z_i=\{n|\braket{\phi_i}{U^n}{\psi_0}=0\}$ is semi-linear. Furthermore, we see that $Z=\cap Z_i$ is also semi-linear. Thus, the infiniteness of $Z$ in Theorem~\ref{the:main} actually implies that it contains at least one arithmetic progression.

There is still a gap between the existence of the arithmetic progression in Theorem~\ref{the:main} and its algorithmic construction. Here we further present an algorithm to find $p$ and $r$ such that $PU^{pn+r}\ket{\psi_0}=0$ for all $n=0,1,\cdots$. Of course we should assume that all operators and states are represented by matrices and vectors of rational complex numbers.

\emph{Finding number $p$:} We can algorithmically find a positive integer $p$ satisfying the following condition: \begin{itemize}\item for any two eigenvalues $\lambda$ and $\mu$ of $U$, $(\lambda/\mu)^p=1$ provided $(\lambda/\mu)^n=1$ for some integer $n$.\end{itemize} Indeed, it suffices to find the smallest positive integer $n$ satisfying $(\lambda/\mu)^n=1$ for each fixed pair of $\lambda, \mu$, and then $p$ can be chosen as the least common multiple of all these $n$. We note that all roots of the characteristic polynomial $f(x)$ of $U\otimes U^\dagger$ are exactly all quotients $\lambda/\mu$ of two eigenvalues of $U$. Moreover, for each quotient $\lambda/\mu$, if $n$ is the smallest positive integer number satisfying $(\lambda/\mu)^n=1$, then $\lambda/\mu$ should be a root of the $n$th cyclotomic polynomial $\Phi_n(x)$, and $\Phi_n(x)$ should be a divisor of $f(x)$ since $\Phi_n(x)$ is irreducible in the field of rational numbers. Therefore, all of such $n$ can be obtained by checking whether or not $\Phi_n(x)|f(x)$.

Moreover, we prove that the number $p$ enjoys an property: for any subspace $K$ of $\hs$, $U^pK=K$ provided $U^nK=K$ for some integer $n$. We observe that $U^nK=K$ if and only if a set of eigenvectors of $U^n$ forms a basis of $K$. From this observation, it suffices to prove that any eigenvector of $U^n$ is an eigenvector of $U^p$. More generally, we show that any eigenspace $E$ of $U^n$ is included in some eigenspace of $U^p$.
We note that all eigenvectors of $U$ are eigenvectors of $U^n$, so we can choose a set of eigenvectors of $U$ to form a basis $B$ of $E$. Consider any two of these vectors, written as $\ket{\psi}$ and $\ket{\phi}$, and we write $\lambda$ and $\mu$, respectively, for the corresponding eigenvalues of $U$. Then we have $(\lambda/\mu)^{n}=1$, and according to our choice of $p$, $(\lambda/\mu)^p=1$. So $\ket{\psi}$ and $\ket{\phi}$ are in the same eigenspace of $U^p$. As these two states are arbitrarily chosen, it implies that all of the vectors in $B$ are in the same eigenspace of $U^p$. Thus $E$ is included in it.

\emph{Finding number $r$:} Let $K=\{\ket{\psi}|P\ket{\psi}=0\}$ be the kernel space of $P$. For any integer $q$, we write $K_q$ for the maximal subspace of $K$ satisfying $U^qK_q=K_q$. Then $K_q$ can be calculated by the iteration $K_q\leftarrow K_q\cap U^qK_q$, putting $K_q\leftarrow K$ initially. On the other hand, we show that \begin{equation}\label{equ:Kq}
K_q=\{\ket{\psi}\in K|U^{nq}\ket{\psi}\in K\ {\rm for\ all\ integer}\ n\geq 0\}.
\end{equation} First, for any state $\ket{\psi}\in K_q$, one can easily verify from the definition of $K_q$ that $U^{nq}\ket{\psi}\in K_q\subseteq K$ for all $n$. Secondly, if some state $\ket{\psi}$ satisfies $U^{nq}\ket{\psi}\in K$ for all $n$, then we consider the subspace of $K$: $K'=\mathrm{span}\{U^{nq}\ket{\psi}|n=0,1,\cdots\}$. We have $U^qK'=K'$, and thus $\ket{\psi}\in K'\subseteq K_q$ from the maximality of $K_q$. Therefore, Eq.~(\ref{equ:Kq}) holds.

To make $U^{pn+r}\ket{\psi_0}=0$ for all $n\geq 0$, it suffices to calculate $K_p$ and then find $r$ from $\{0,1,\cdots,p-1\}$ such that $U^r\ket{\psi_0}\in K_p$. Now we only need to prove the following claim: \begin{itemize}\item Whenever there exists an arithmetic progression $\{an+b|n=0,1,\cdots\}$ in $Z$, the number $r$ can be found as above.\end{itemize}  In fact, by Eq. (\ref{equ:Kq}), $\{an+b|n=0,1,\cdots\}\subseteq Z$ means that $U^b\ket{\psi_0}\in K_a$. We note that $U^aK_a=K_a$ implies $U^pK_a=K_a$ by the property of $p$ stated above. Thus, $K_a\subseteq K_p$ due to the maximality of $K_p$. So we have $U^b\ket{\psi_0}\in K_p$. If we put $r=b-cp\in\{0,1,\cdots,p-1\}$ as the remainder of $b$ divided by $p$, then $U^r\ket{\psi_0}\in U^{-cp}K_p=K_p$. So $r$ can be obtained in the algorithm. This completes the proof of Theorem~\ref{the:main}.

\subsection{Construction of the Debugging Protocols}
Now we can construct all debugging protocols $(P,S)$ for a given process using the proof of Theorem~\ref{the:main}. A necessary and sufficient condition of error detectors $P$ is that $PU^{pn+r}\ket{\psi_0}=0\ (n\geq 0)$ for the integer $p$ and some $r\in\{0,1,\cdots,p-1\}$. So the construction of $(P,S)$ is achieved in four steps:\begin{enumerate}\item Compute the number $p$ from the given unitary operator $U$. An algorithm for finding $p$ was already presented in the proof of Theorem~\ref{the:main}. \item Arbitrarily choose a number $r\in\{0,1,\cdots,p-1\}$, and compute the subspace $$V_r={\rm span}\{U^{pn+r}\ket{\psi_0}|n=0,1,\cdots,d-1\},$$ where $d$ is the dimension of the system. \item $P$ can be chosen as any projection operator satisfying $PV_r=0$. In particular, we choose it as the one with image space $V_r^\perp$, since it is of the maximal rank and thus can detect as many as possible errors. \item $S$ is constructed as the periodic function that specifies the arithmetic progression $\{pn+r|n=0,1,\cdots\}$.
\end{enumerate}

As an instance, we show how the above procedure can be used to construct a debugging protocol for the quantum search process in Subsec.~\ref{sec:proto}-A. The computational process of quantum search can be formalized in our model: the Hilbert space $\hs$ is of dimension $N=2^n$, the initial state is
$\ket{\psi_0}=\sum_{k=0}^{N-1}\ket{k}/\sqrt{N}$ and the unitary transformation is
$$G=(2\op{\psi_0}{\psi_0}-I_2^{\otimes n})(I_2^{\otimes n}-\op{x}{x}),$$
where $x\in\{0,1,\cdots,N-1\}$ is a given integer. Then the number $p$, number $r$, and projection operator $P$ are determined as follows:
\begin{enumerate}\item To obtain the number $p$, we calculate the characteristic polynomial of $G$ that is
$$(\lambda-1)^{N-2}(\lambda^2+2(1-2/N)\lambda+1).$$
We only consider the case of $N>4$. It is easy to verify that for any two eigenvalues $\lambda$, $\mu$ of $G$,
if $(\lambda/\mu)^n=1$ for some $n$ then $\lambda=\mu$. So we have $p=1$.
 \item Now $r\in\{0,1,\cdots,p-1\}$ can only be $0$ because $p=1$. Then
 \begin{equation*}\begin{split}
 V_0&={\rm span}\{U^n\ket{\psi_0}|n=0,1,\cdots,N-1\}\\
 &={\rm span}\{\ket{x},\ket{\xi}\},
 \end{split}\end{equation*}
 where $\ket{\xi}=\sum_{k\neq x}\ket{k}/\sqrt{N-1}$.
 \item We choose $P=I_2^{\otimes n}-\op{x}{x}-\op{\xi}{\xi}$ to make the condition $PV_0=0$ be satisfied. \end{enumerate}
As $p=1$ and $r=0$, $P$ is applied immediately after each action of $G$. We see that this protocol constructed by the procedure presented in this subsection is exactly that given in Subsection \ref{sec:proto}-A.

\section{Conclusion}\label{sec:conclu}
In this paper, we proposed a scheme for debugging a quantum process, in which quantum measurements are used to monitor the system without disturbances on its behaviour. We discovered a procedure to construct all debugging protocols in this scheme for quantum processes with time independent Hamiltonians. However, the problem of debugging quantum processes is still open for the case of time dependent Hamiltonians.

\section*{Acknowledgement}
We are grateful to Runyao Duan, Yuan Feng and Nengkun Yu for useful discussions. This work was
partly supported by the Australian Research Council (Grant No: DP110103473 and DP130102764).


\begin{thebibliography}{99}
\bibitem{Sho96} P. W. Shor, in Proc. 37th Annual Symposium on Foundations of Computer Science, 56-65 (IEEE Press, Los Alamitos, 1996).
\bibitem{Kit97} A. Y. Kitaev, Russ. Math. Surv. \textbf{52}, 1191 (1997).
\bibitem{Sho95} P. W. Shor, Phys. Rev. A \textbf{52}, 2493 (1995).
\bibitem{Ste96} A. M. Steane, Phys. Rev. Lett. \textbf{77}, 793 (1996).
\bibitem{NC00} M. A. Nielsen and I. L. Chuang, \emph{Quantum Computation and Quantum Information} (Cambridge University Press, 2000).
\bibitem{KLZ98} E. Knill, R. Laflamme, and W. H. Zurek, Science \textbf{279}, 342 (1998).
\bibitem{ZSBS14} J. Zhang, A. M. Souza, F. D. Brandao, and D. Suter, Phys. Rev. Lett. \textbf{112}. 050502 (2014).
\bibitem{Mye79} M. Leucker and C. Schallhart, Journal of Logic and Algebraic Programming, \textbf{78}, 293 (2009); G. J. Myers, \emph{The Art of Software Testing} (John Wiley and Sons, Inc. 1979).
\bibitem{DFY09} R. Y. Duan, Y. Feng, and M. S. Ying, Phys. Rev. Lett. \textbf{103}, 210501 (2009).
\bibitem{BBC+93} C. H. Bennett, G. Brassard, C. Cr\'{e}peau, R. Jozsa, A. Peres, and W. K. Wootters, Phys. Rev. Lett. \textbf{70}, 1895 (1993).
\bibitem{BBP+96} C. H. Bennett, G. Brassard, S. Popescu, B. Schumacher, J. A. Smolin, and W. K. Wootters, Phys. Rev. Lett. \textbf{76}, 722 (1996).
\bibitem{AT12} C. Altafini, and F. Ticozzi, IEEE Transactions on Automatic Control, \textbf{57}, 1898 (2012).
\bibitem{RB01} R. Raussendorf and H. J. Briegel, Phys. Rev. Lett. \textbf{86}, 5188 (2001).
\bibitem{MS77} B. Misra and E. C. G. Sudarshan, J. Math. Phys. \textbf{18}, 756 (1977).
\bibitem{PCZ97} J. F. Poyatos, J. I. Cirac, and P. Zoller, Phys. Rev. Lett. \textbf{78}, 390 (1997); I. L. Chuang and M. A. Nielsen, J. Mod. Opt. \textbf{44}, 2455 (1997).
\bibitem{Gro96} L. Grover, in Proc. 28th Annual ACM Symposium on the Theory of Computing, 212-219 (ACM Press, New York, 1996).
\bibitem{ABN+01} A. Ambainis, E. Bach, A. Nayak, A. Vishwanath, and J. Watrous, in Proc. 33rd Annual ACM Symposium on the Theory of Computing, 37-49 (ACM Press, New York, 2001).
\bibitem{Skolem34} T. Skolem, in Proc. 8th Congress of Scandinavian Mathematicians, 163-188, (Stockholm, 1934).
\bibitem{BJKP05} V. D. Blondel, E. Jeandel, P. Koiran and N. Portier, SIAM J. Comput. \textbf{34}, 1464 (2005).
\bibitem{EMG12} J. Eisert, M. P. M\"{u}ller and C. Gogolin, Phys. Rev. Lett. \textbf{108}, 260501 (2012).
\bibitem{HHHK05} V. Halava, T. Harju, M. Hirvensalo and J. Karhumaki, TUCS Technical Report, 683 (2005).
\bibitem{Lech52} C. Lech, Ark. Mat. \textbf{2}, 417-421 (1953).
\end{thebibliography}
\end{document}